\documentstyle[prl,aps,epsfig,preprint,array]{revtex}

\def\Aut{|A\widetilde{\uparrow}\rangle}
\def\Adt{|A\widetilde{\downarrow}\rangle}

\def\Bu{|B\!\uparrow\rangle}
\def\Bd{|B\!\downarrow\rangle}
\def\But{|B\widetilde{\uparrow}\rangle}
\def\Bdt{|B\widetilde{\downarrow}\rangle}

\begin{document}


\title{Entanglement Concentration Using Quantum Statistics}

\author{N. Paunkovi\'{c}$^{1}$, Y. Omar$^{1}$, S. Bose$^{1}$ and V.
Vedral$^{2}$}

\address{$^{1}$ Centre for Quantum Computation, Clarendon Laboratory,
University of Oxford, Oxford OX1 3PU, United Kingdom\\
$^{2}$ Optics Section, The Blackett Laboratory, Imperial College,
London SW7 2BZ, United Kingdom \\}

\date{30 November 2001}

\maketitle


\begin{abstract}
We propose an entanglement concentration scheme which uses only
the effects of quantum statistics of indistinguishable particles.
This establishes the fact that useful quantum information
processing can be accomplished by quantum statistics alone. Due
to the basis independence of statistical effects, our protocol
requires less knowledge of the initial state than most
entanglement concentration schemes. Moreover, no explicit
controlled operation is required at any stage.
\end{abstract}



Quantum statistics has led to a number of interesting phenomena
such as ferromagnetism, superconductivity and superfluidity.
However, its effects have never been used to process information.
Recently, we have shown that quantum statistics can lead to an
\textit{effective} interaction between internal and external
degrees of freedom of particles \cite{omar}. This was demonstrated
in the context of entanglement transfer from spin to path without
an explicit conditional operation between these degrees of
freedom. It was also shown that the extent of this transfer
depended on the bosonic or fermionic nature of the particles
\cite{omar}. Going even further in our exploration of the effects
of quantum statistics in information theory, we present an
entanglement concentration protocol \cite{gisin,bennett1} based
on these effects. This establishes the fact that quantum
statistics {\em alone} (without any other explicit interaction
between the relevant degrees of freedom of the particles
involved) is sufficient for useful quantum information processing.

The best performances in quantum communication and computation
processes are normally achieved using a pair of pure maximally
entangled systems (particles) shared between distant parties. But
the inevitable influence of the environment during the
distribution of the entangled pairs reduces the amount of shared
entanglement. As entanglement cannot be increased by local
operations and classical communication \cite{vlatko}, the only
option is to concentrate it locally from a larger to a smaller
number of pairs
\cite{gisin,bennett1,bennett2,deutsch,bose99,horodecki,thew01,kwiat,zeilinger}.
In the case of mixed states we call these protocols {\it
entanglement purification}, while for pure states we refer to them
as {\it entanglement concentration} \cite{bennett1}.

In a setting similar to previous entanglement concentration
protocols, we consider the action of specific local operations on
two pairs of entangled systems, driving them with some probability
into one pair of more entangled systems.

\begin{figure}[ht]
\begin{center}
\epsfig{file=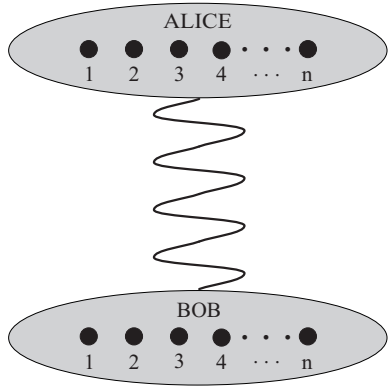, width=2in}
\end{center}
\caption{This figure represents one of the two entangled pairs
that we use for our protocol. Each  pair is composed of two
$n$-particle states entangled between Alice and Bob with at most 1
e-bit of entanglement.} \label{Fig. Initial}
\end{figure}

However, unlike most of the previous schemes, we require our
entangled systems to be composed of $n$ particles (see Fig.
\ref{Fig. Initial} for the representation of a pair of these
systems). We look at the entanglement in the internal degrees of
freedom of the particles, such as the spin in the case of
electrons (fermions) or the polarization in the case of photons
(bosons), which have isomorphic Hilbert spaces. The initial pure
state of each of our two pairs, distributed between two parties
Alice ($A$) and Bob ($B$), is:
\begin{equation}
\label{first} |\phi\rangle^n \equiv \alpha|
\underbrace{\uparrow\uparrow \cdots \uparrow}_{n} \, \rangle_A
|\downarrow\downarrow \cdots
\downarrow\rangle_B+\beta|\downarrow\downarrow \cdots
\downarrow\rangle_A|\uparrow\uparrow \cdots \uparrow\rangle_B
\end{equation}
with $|\alpha|^2+|\beta|^2=1$ and where, for instance, we have:
\begin{equation}
| \uparrow\uparrow \cdots \uparrow \, \rangle_A =
\hat{a}^+_{A1\uparrow}\hat{a}^+_{A2\uparrow} \cdots
\hat{a}^+_{An\uparrow}|0\rangle \equiv \Aut^n.
\end{equation}
Here $\hat{a}^+$ are creation operators that ensure the ordering
of the multi-particle state, as the usual commutation and
anti-commutation rules apply for bosons and fermions respectively.
For sake of compactness, we will rewrite equation (\ref{first})
as:
\begin{equation}
\label{firstcompact} |\phi\rangle^n= \left(
\alpha\Aut^n\Bdt^n+\beta\Adt^n\But^n \right),
\end{equation}
where the tilde over the arrow reminds us of the string of spins
(or polarizations).

\begin{figure}[ht]
\begin{center}
\epsfig{file=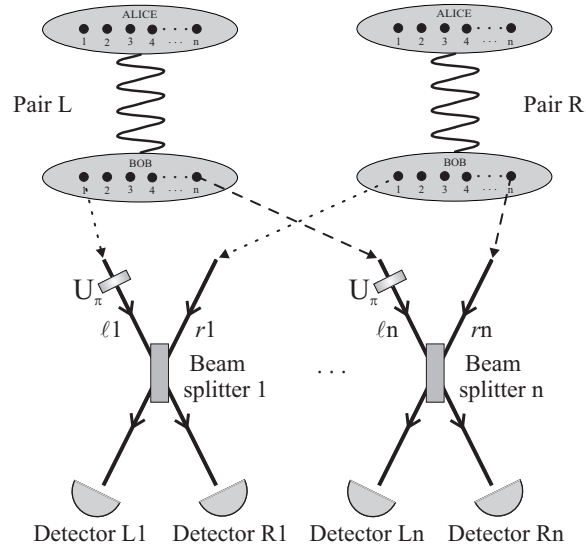, width=3in}
\end{center}
\caption{This figure represents the setup for our entanglement
concentration protocol using quantum statistics. Initially, Alice
and Bob share two pairs of entangled systems, L and R. Each
entangled pair is composed of two $n$-particle states. One of the
parties, say Bob, can then convert these pairs into a more
entangled pair by performing a set of local operations using only
standard 50/50 beam splitters, path measurements and one-way
classical communication with Alice.} \label{Fig. Setup}
\end{figure}

Let us now label one pair $L$ and the other $R$ (as in left and
right -- see Fig. \ref{Fig. Setup}). Then, our total initial state
is:
\begin{equation}
\label{total} |\varphi\rangle^n \equiv
|\phi\rangle^n_L\otimes|\phi\rangle^n_R,
\end{equation}
where $|\phi\rangle^n_L$ is given by equation (\ref{first})
written for pair $L$, and similarly for $R$.

Using equation (\ref{firstcompact}), we can rewrite the total
state (\ref{total}) as:
\begin{eqnarray} \label{fi}
|\varphi\rangle^n & = &
\alpha^2\Aut^n_{L}\Aut^n_{R}\Bdt^n_{L}\Bdt^n_{R}
+\beta^2\Adt^n_{L}\Adt^n_{R}\But^n_{L}\But^n_{R} \nonumber \\
& + &\alpha\beta \left(\Aut^n_{L}\Adt^n_{R}\Bdt^n_{L}\But^n_{R}
+\Adt^n_{L}\Aut^n_{R}\But^n_{L}\Bdt^n_{R} \right).
\end{eqnarray}

Our protocol consists of bringing each of Bob's particles into a
50/50 beam splitter (particles $i$ from the left and the right
systems go into beam splitter $i$, for $ i = 1, 2, \ldots, n $)
and then making a specific measurement on the output particles
using detectors $Li$ and $Ri$ (see Fig. \ref{Fig. Setup}). Before
the particles from the left fall into the beam splitter, we apply
to each of them a specific unitary transformation $U^i_\pi$, that
flips their spin or polarization. This is done in the arm $\ell
i$ of each beam splitter and the transformation is defined for
every $i$:
\begin{equation}
U^i_\pi: \left\{
\begin{array}{l}
|B\uparrow\rangle_{\ell i} \rightarrow |B\downarrow\rangle_{\ell i} \\
|B\downarrow\rangle_{\ell i} \rightarrow |B\uparrow\rangle_{\ell
i}
\end{array}
\right. .
\end{equation}
This flipping is applied in order to align some spins (or
polarizations) in appropriate terms so that we can exploit the
statistical effects later on.

The total state then becomes:
\begin{eqnarray} \label{fi0}
|\varphi_0\rangle^n & \equiv & \left( U^1_\pi \otimes U^2_\pi
\otimes \cdots  \otimes U^n_\pi \right) |\varphi\rangle^n \nonumber \\
& = & \alpha^2\Aut^n_{L}\Aut^n_{R}\But^n_{L}\Bdt^n_{R}
+\beta^2\Adt^n_{L}\Adt^n_{R}\Bdt^n_{L}\But^n_{R} \nonumber \\
& + & \alpha\beta \left(\Aut^n_{L}\Adt^n_{R}\But^n_{L}\But^n_{R}
+\Adt^n_{L}\Aut^n_{R}\Bdt^n_{L}\Bdt^n_{R} \right).
\end{eqnarray}

First, we present the entanglement concentration protocol for
fermions (e.g. electrons) and then the counterpart protocol for
bosons (e.g. photons).

{\bf \textit{Fermions}} -- After passing through the beam
splitter, Bob performs path measurement on the first pair of
particles coming out of the beam splitter using detectors $L1$ and
$R1$ (see Fig. \ref{Fig. Setup}). For simplicity of presentation,
we first assume that these detectors do not absorb the particles
and do not disturb their internal degrees of freedom (such
detectors are indeed available for electrons and atoms and can
have interesting applications in deterministically entangling
independent particles using indistinguishability and feedback
\cite{bose01}). However, as we will show later on, this is not a
necessary requirement for success of our protocol. If the
particles (which we assumed to be fermions) bunch, we discard
them. Otherwise, in the case of anti-bunching, we get the pure
state:
\begin{eqnarray} \label{fi1}
|\varphi_1\rangle^n & = & N_1 \left[ \frac{1}{\sqrt2} \right.
\left(
\alpha^2\Aut^n_{L}\Aut^n_{R}|B\,triplet\rangle_{L1R1}\But^{n-1}_{L}\Bdt^{n-1}_{R}
\right.
\nonumber \\
& + & \beta^2 \left.
\Adt^n_{L}\Adt^n_{R}|B\,triplet\rangle_{L1R1}\Bdt^{n-1}_{L}\But^{n-1}_{R}
\right)
\nonumber \\
& + & \alpha\beta \left.
\left(\Aut^n_{L}\Adt^n_{R}\But^n_{L}\But^n_{R}
+\Adt^n_{L}\Aut^n_{R}\Bdt^n_{L}\Bdt^n_{R} \right) \right],
\end{eqnarray}
where
\begin{equation}
N_1=\left[\frac{|\alpha|^4}{2}+\frac{|\beta|^4}{2} +
2|\alpha\beta|^2\right]^{-\frac{1}{2}},
\end{equation}
and the state
\begin{equation}
|B\,triplet\rangle_{L1R1} \equiv \frac{1}{\sqrt2} \left(
|B\uparrow\rangle_{L1}|B\downarrow\rangle_{R1}+
|B\downarrow\rangle_{L1}|B\uparrow\rangle_{R1} \right)
\end{equation}
is a triplet state, the anti-bunching result of Bob's measurement
with detectors $L1$ and $R1$ on the first pair of particles.

The probability $p_1$ of having anti-bunching and thus getting
state $|\varphi_1\rangle^{n}$ (in an ideal measurement) is:
\begin{equation}
\label{prob1}
p_1=\frac{|\alpha|^4}{2}+\frac{|\beta|^4}{2}+2|\alpha\beta|^2=N_1^{-2},
\end{equation}
because the first two terms have a probability $\frac{1}{2}$ of
anti-bunching, while the other two have probability $1$ (we
normalize these probabilities by  $|\alpha|^2$, $|\beta|^2$ and
$|\alpha\beta|^2$, the probabilities of obtaining the
corresponding results). Thus, the state $|\varphi_1\rangle$
differs from $|\varphi_0\rangle$ in the first two terms (we
"normalize" their amplitudes $\alpha^2$ and $\beta^2$ by the
factor $\frac{1}{\sqrt2}$).

After obtaining the anti-bunching result, we can now let the
second pair of Bob's particles ($\ell 2$ and $r2$) pass through
another 50/50 beam splitter and perform the same measurement,
discarding again the bunching results.

After repeating the same procedure $n$ times, we get the state:
\begin{eqnarray} \label{fin}
|\varphi_n\rangle^n & = & N_n\ \left[
\frac{\alpha^2}{2^\frac{n}{2}}
\Aut^n_{L}\Aut^n_{R}|B\,triplet\rangle^{\otimes{n}}
+\frac{\beta^2}{2^\frac{n}{2}}
\Adt^n_{L}\Adt^n_{R}|B\,triplet\rangle^{\otimes{n}} \right. \nonumber \\
& + & \alpha\beta \left. \left(
\Aut^n_{L}\Adt^n_{R}\But^n_{L}\But^n_{R}+
\Adt^n_{L}\Aut^n_{R}\Bdt^n_{L}\Bdt^n_{R} \right)\  \right],
\end{eqnarray}
with
\begin{equation}
N_n=\left[\frac{|\alpha|^4}{2^n}+\frac{|\beta|^4}{2^n}
+2|\alpha\beta|^2\right]^{-\frac{1}{2}}.
\end{equation}

The probability of getting $|\varphi_n\rangle$ from $|\varphi_{n-1}\rangle$
is:
\begin{equation}
\label{probn} p_n=\left( \frac{N_{n-1}}{N_n} \right)^2.
\end{equation}

So, the total probability for getting $|\varphi_n\rangle^n$ from
$|\varphi_0\rangle^n$ is:
\begin{equation}
\tilde p_n=\prod_{i=1}^n
p_i=N_n^{-2}\stackrel{(n\rightarrow\infty)}{\longrightarrow}2|\alpha\beta|^2\equiv\tilde p.
\end{equation}

After applying the same procedure "infinitely many times" (to a
"infinitely large state"), we obtain the state:

\begin{eqnarray}
|\varphi_\infty\rangle & \equiv &
\lim_{n\rightarrow\infty}|\varphi_n\rangle^n \nonumber \\
& = & \lim_{n\rightarrow\infty} \frac{1}{\sqrt{2}}
\left(\Aut^n_{L}\Adt^n_{R}\But^n_{L}\But^n_{R}+\Adt^n_{L}\Aut^n_{R}\Bdt^n_{L}\Bdt^n_{R}
\right).
\end{eqnarray}

The total probability is then:
\begin{equation}
\label{prob} p = \frac{\tilde p}{2} = |\alpha\beta|^2.
\end{equation}

Now note that even if we had, instead of non-absorbing, any kind
of path detectors, our protocol would still work assuming we
perform a path measurement on $n-1$ particles. We would then have
absorbed all these $n-1$ particles, and their state would be
replaced by the vacuum state $|0\rangle$. In the
$n\rightarrow\infty$ limit this will factorize out to give the
final maximally entangled state:
\begin{equation}
|\varphi_\infty\rangle= \lim_{n\rightarrow\infty}
\frac{1}{\sqrt{2}}
\left(\Aut^n_{L}\Adt^n_{R}\Bu_{L}\Bu_{R}+\Adt^n_{L}\Aut^n_{R}\Bd_{L}\Bd_{R}
\right).
\end{equation}

The probability $p$ given by equation (\ref{prob}) provides us
with a measure of efficiency of our entanglement concentration
protocol. It is the amount of entanglement in e-bits that we can
extract from a single pair of entangled particles in the initial
state $|\phi \rangle^n$ given by equation (\ref{firstcompact}),
in the limiting case:
\begin{equation}
\label{efficiency} E^\infty(|\phi\rangle^{n})= p\,,
(n\rightarrow\infty).
\end{equation}

{\bf \textit{Bosons}} -- In this case, the procedure is almost the
same as for fermions, but this time we discard the anti-bunching
results and keep the bunching ones.

After Bob measures the first pair of particles, we obtain a state
similar to (\ref{fi1}), but where the particles will now be
either both in $L1$ or either both in $R1$. In other words,
instead of the triplet state $|B\,triplet\rangle_{L1R1}$, we have
the term $\frac{1}{\sqrt{2}}(\Bu_{L1}\Bd_{L1}+\Bu_{R1}\Bd_{R1})$,
while instead of $\But^n_{L}\But^n_{R}$ we have
$\frac{1}{\sqrt{2}}(\But_{L1}\But_{L1}+\But_{R1}\But_{R1})\But^{n-1}_{L}\But^{n-1}_{R}$,
and similarly for the last term in (\ref{fi1}).

The probability of getting this state is again given by
(\ref{prob1}), $p_1=N_1^{-2}$.

By reversing the selection of the path measurements performed by
Bob (i.e. by now discarding the anti-bunching instead of the
bunching results), we have established a symmetry between the
bosonic and the fermionic protocols. The general state
(\ref{fin}) and its probability (\ref{probn}) are in fact the
"same" (isomorphic), as well as the total probability
(\ref{prob}), $p=|\alpha\beta|^2$, and the efficiency
(\ref{efficiency}):
\begin{equation}
E^\infty(|\phi\rangle^{n})= p\,, (n\rightarrow\infty).
\end{equation}

Note that for both protocols (for fermions and for bosons) the
unitary transformation $U_\pi \equiv \bigotimes^{n}_{i=1} U^i_\pi$
is crucial. Since states $| \varphi \rangle^n$ and $| \varphi_0
\rangle^n = U_\pi | \varphi \rangle^n$ -- given by equations
(\ref{fi}) and (\ref{fi0}) respectively -- are isomorphic, one
might expect the results to be the same regardless of $U_\pi$
being applied or not. This turns out to be wrong, as applying the
protocol directly to $| \varphi \rangle^n$ will yield the
following state:
\begin{eqnarray}
|\varphi_\infty\rangle & \equiv &
\lim_{n\rightarrow\infty}|\varphi_n\rangle \nonumber \\
& = & \lim_{n\rightarrow\infty}
\frac{1}{\sqrt{|\alpha|^4+|\beta|^4}}
\left(\alpha^2\Aut^n_{L}\Aut^n_{R}\Bdt^n_{L}\Bdt^n_{R}
+\beta^2\Adt^n_{L}\Adt^n_{R}\But^n_{L}\But^n_{R} \right),
\end{eqnarray}
in general less entangled than the initial one.

In the protocols we have presented here, local operations are
performed on one side only (Bob). Classical communication comes
about only once, when we have one way communication from Bob to
Alice at the end of the whole scheme. Of course, we can slightly
change our protocols by allowing Alice to apply the same
procedure on her side. This would require two way communication
and would have half the time complexity of the protocol, but the
amount of entanglement distilled would be the same. Also, note
that both Alice and Bob could perform the operations on the $n-1$
particles, each one on their side, using either one beam splitter
sequentially or $n-1$ beam splitters in parallel. The latter case
has a lower time complexity than the former, but requires more
resources (higher space complexity).

The efficiency $|\alpha \beta|^2$ of our protocol happens to be
lower than both the efficiency $2|\alpha|^2$ of the procrustean
method and the efficiency
$-|\alpha|^2\mbox{Log}|\alpha|^2-|\beta|^2\mbox{Log}|\beta|^2$ of
the standard asymptotic entanglement concentration procedure
\cite{bennett1}. However, our target is {\em not} to propose a
more efficient {\em alternative} entanglement concentration, but
to explicitly demonstrate that it is {\em possible} to do
entanglement concentration by using {\em only} the effects of
particle statistics, without resort to an explicit controlled
operation between spins or between spin and path. In all physical
implementations of entanglement concentration or purification
protocols to date \cite{kwiat,zeilinger}, a polarization beam
splitter or a polarization dependent filter (which accomplish a
controlled operation on path conditional on spin) have been used.

It is important to note here that our mechanism will work even if
the basis states $| \! \! \uparrow\rangle$ and $| \! \!
\downarrow\rangle$, were rotated in a plane. This is because the
only unitary rotation we use is $U_{\pi}$, which will flip all
spins in one plane. The rest of the protocol uses particle
statistics, which is basis independent. On the other hand, for
standard entanglement concentration protocols
\cite{gisin,bennett1}, it is necessary to know the basis.

In this paper we have presented an entanglement concentration
scheme which uses \textit{only} the effects of quantum
statistics. Although the efficiency of the protocol is the same
for both fermions and bosons, the protocol itself is slightly
different depending on the nature of the particles. This brings
forth a fundamental difference between these two types of
particles in terms of their information processing power. Recent
experiments such as \cite{kwiat2,yamamoto} suggest that it would
be possible to test our results in the near future. It would also
be interesting to investigate the possibility of entanglement
distillation of mixed states using {\em solely} statistical
effects. Future work will comprise of investigating more
extensive quantum information processing procedures.

N.P. thanks Elsag S.p.A. for financial support. Y.O. acknowledges
support from Funda\c{c}\~{a}o para a Ci\^{e}ncia e a Tecnologia
from Portugal. V.V. acknowledges support from Hewlett-Packard
company, EPSRC and the European Union project EQUIP.


\begin{thebibliography}{20}
%
\bibitem{omar} Y. Omar, N. Paunkovi\'c, S. Bose and V. Vedral, quant-ph/0105120 (2001).
%
\bibitem{gisin} N. Gisin, Phys. Lett. A {\bf 210}, 151 (1996).
%
\bibitem{bennett1} C. H. Bennett, H. J. Bernstein, S. Popescu, and B. Schumacher, Phys. Rev. A
 {\bf 53}, 2046 (1996).
%
\bibitem{vlatko} M. Plenio and V. Vedral, Cont. Phys. {\bf 39}, 431 (1998).
%
\bibitem{bennett2} C.H. Bennett, D. P. DiVicenzo, J. A. Smolin, and W. K. Wootters, Phys. Rev. A
 {\bf 54}, 3824 (1996).
%
\bibitem{deutsch} D. Deutsch \textit{et al.}, Phys. Rev. Lett. {\bf 77}, 2818 (1996).
%
\bibitem{bose99} S. Bose, V. Vedral and P. L. Knight, Phys. Rev. A {\bf 60}, 194
(1999); L. Hardy and D. D. Song, Phys. Rev. A {\bf 62}, 052315
(2000).
%
\bibitem{horodecki} M. Horodecki, P. Horodecki and R. Horodecki, Phys. Rev. Lett. {\bf
78}, 574 (1997).
%
\bibitem{thew01}
R. T. Thew and W. J. Munro, Phys. Rev. A {\bf 63}, 030302 (2001).
%
\bibitem{kwiat} P. G. Kwiat, S. Barraza-Lopez, A. Stefanov and N. Gisin, Nature
{\bf 409}, 1014 (2001).
%
\bibitem{zeilinger} J. W. Pan, C. Simon, \v{C}. Brukner and A. Zeilinger, Nature
 {\bf 410}, 1067 (2001).
%
\bibitem{bose01} S. Bose and D. Home, Phys. Rev. Lett. (to appear), quant-ph/0101093.
%
\bibitem{kwiat2} T. J. Herzog, P.G. Kwiat, H. Weinfurter and A.
Zeilinger, Phys. Rev. Lett. {\bf 75}, 3034 (1995).
%
\bibitem{yamamoto} R.C. Liu, B. Odom, Y. Yamamoto and S. Tarucha, Nature {\bf 391}, 263 (1998).
%
\end{thebibliography}
\end{document}